\documentclass[
   ,final          
  ]
  {aipproc}

\layoutstyle{6x9}

\begin{document}

\title{Diffuse Neutron Scattering Study of Relaxor Ferroelectric (1-$x$)Pb(Zn$_{1/3}$Nb$_{2/3}$)O$_{3}$-$x$PbTiO$_{3}$ (PZN-$x$PT)}

\author{D. La-Orauttapong}{
  address={Department of Physics, Lehigh University, Bethlehem, Pennsylvania 18015-3182}
}
\author{J. Toulouse}{
  address={Department of Physics, Lehigh University, Bethlehem, Pennsylvania 18015-3182}
}

\author{Z.-G. Ye}{
  address={Department of Chemistry, Simon Fraser University, Burnaby, British Columbia,
Canada V5A 1S6}
}

\author{R. Erwin}{
  address={NIST Center for Neutron Research, NIST, Gaithersburg, Maryland 20899-8562}
}

\author{J.L. Robertson}{
  address={Oak Ridge National Laboratory, Solid State Division, Oak Ridge, Tennessee
37831-6393}
}
\author{W. Chen}{
  address={Department of Chemistry, Simon Fraser University, Burnaby, British Columbia,
Canada V5A 1S6}
}

\begin{abstract}
Diffuse neutron scattering is a valuable tool to obtain
information about the size and orientation  of the polar nanoregions that
are a characteristic feature of relaxor ferroelectrics. In this paper, we
present  new diffuse scattering results obtained on Pb(Zn$_{1/3}$Nb$_{2/3}$)O%
$_{3}$ (PZN for short) and  (1-$x$)Pb(Zn$_{1/3}$Nb$_{2/3}$)O$_{3}$-$x$PbTiO$%
_{3}$ (PZN-$x$PT) single crystals (with $x$=4.5 and 9\%),  around various
Bragg reflections and along three symmetry directions in the [100]-[011]
zone.  Diffuse scattering is observed around reflections with mixed indices,
(100), (011) and (300), and along  transverse and diagonal directions only.
No diffuse scattering is found in longitudinal scans.  The diffuse
scattering peaks can be fitted well with a Lorentzian function, from which
a correlation length is extracted. The correlation length
increases with decreasing temperatures  down to the transition at $T_c$,
first following a Curie-Weiss law, then departing from it and becoming flat
at very  low temperatures. These results are interpreted in terms of three
temperature regions: 1) dynamic polarization  fluctuations (i.e. with a
finite lifetime) at high temperatures, 2) static polarization
reorientations  (condensation of polar nanoregions) that can still reorient as
a unit (relaxor behavior) at intermediate temperatures  and 3) orientational
freezing of the polar nanoregions with random strain fields in pure PZN or a
structural phase  transition in PZN-$x$PT at low temperatures. The addition
of PT leads to a broadening of the diffuse scattering along  the diagonal
([111]) relative to the transverse ([100]) direction, indicating a change in
the orientation of the polar  regions. Also, with the addition of PT, the
polar nanoregions condense at a higher temperature above $T_c$.
\end{abstract}

\maketitle


\section{Introduction}

The single-crystal solid solution of the relaxor Pb(Zn$_{1/3}$Nb$_{2/3}$)O$%
_{3}$ and ferroelectric PbTiO$_{3}$, known as PZN-$x$PT is being considered
as a promising candidate for the next generation material for
electromechanical transducers \cite{Park_etal_1804_1997}. This material has
cubic perovskite structure at high temperatures but becomes slightly distorted at
lower temperatures, where a transition to ferroelectric/relaxor phase takes
place. Close to pure PZN (for PT contents below 9\%), the low temperature phase
is rhombohedral (R3m) and the material is a relaxor. For higher PT contents,
the low temperature phase of the material becomes tetragonal (P4mm) and the relaxor 
character gradually vanishes. The boundary between the rhombohedral and tetragonal 
phases is called the ``morphotropic phase boundary'' (MPB), near which the crystals show
remarkably large piezoelectric coefficients \cite{Park_etal_1804_1997,Kuwata_etal_1298_1982}. 
In earlier studies of a rhombohedral PZN-8\%PT crystal in [001] bias electric fields, Park and Shrout 
\cite{Park_etal_1804_1997} proposed that the origin of the large strain
values observed is a rhombohedral-to-tetragonal phase transition. Later,
Liu {\it et al.}\cite{Liu_etal_2810_1999} reported a similar behavior in
PZN-4.5\%PT. Very recently, Noheda {\it et al.}\cite{Noheda_etal_field} have also observed 
a field-induced long-range tetragonal phase in both PZN-4.5\%PT and 
PZN-8\%PT using X-ray investigation, which is in agreement with the
model proposed by Park and Shrout. Recently, the MPB has been 
characterized on single crystal samples by high resolution x-ray
measurements \cite{Cox_etal_APL_400_2001,La-Orauttapong_etal}. An
orthorhombic phase (space group Pm) has been found in a narrow concentration
range ($8\%<x<11\%$) with near vertical phase boundaries between the
tetragonal and rhombohedral phases. It can be described as a ``matching''
phase between the rhombohedral and tetragonal phase, which allows for a very
easy reorientation of the polarization vector \cite{Fu and
Cohen_Nature_281_2000}. Recently, a model has been proposed that
relates the structural features of the lead relaxor systems to their unusual
polarization properties \cite{Vanderbilt and Cohen_PRB_094108_2001}.

In addition to the above structural features, there exists another aspect of
the polarization properties of relaxor systems, whose precise connection to
these structural features has not been established yet. \ It is now a well 
recognized fact that, as relaxors are cooled from high temperature, 
the structural changes are preceded by appearance and growth of polar nanoregions. 
The first indirect observation of such regions came from birefringence measurements by
Burns and Dacol on PMN and PZN \cite{Burn_etal_853_1983}. These measurements
revealed deviations of the birefringence $\Delta n(T)$ ($=n_{\parallel
}-n_{\perp }$) from a linear temperature dependence, at temperature $T_{d}$, 
far above the temperature range in which PZN displays
the relaxor behavior.\ These deviations were interpreted as
marking the appearance of the polar regions. Since then, several other experimental 
observations have been reported that support the appearance of the polar regions in lead relaxors
below 650 K for PMN and below 750 K for PZN. The most direct ones have come
from measurements of diffuse X-ray or neutron scattering \cite{Mathan
etal_8159_1991,You and Zhang_PRL_3950_1997}. \ Recently, we have
reported diffuse neutron scattering results on a PZN single crystal that
clearly indicate the development of short range order (a few unit cells) below
approximately $T_{d}=750$K \cite{La-Orauttapong_etal_PRB}. From the width of
the diffuse scattering peak measured around the (011) Bragg reflection, we
have been able to determine a correlation length or the size of the regions
as a function of temperature (550 K-295 K). At the beginning of this range the temperature
dependence of the correlation length is consistent with the Curie-Weiss
dependence of the dielectric constant, which indicates the dynamic character
of polarization. In this temperature range, therefore, the crystal behaves
as a perfect paraelectric. At about $T^{\ast
}=T_{c}+\delta T$ ($\delta T$=40 K), the deviation from the described above linear
dependence indicates that the orientation of the polar regions becomes
progressively more and more static. \ This progressive freezing is
accompanied by the development of permanent strain fields that cause the
Bragg intensity to increase rapidly (relief of extinction). In the present
paper, we report the results of a diffuse neutron scattering study of
several PZN-$x$PT single crystals with $x$=0, 4.5, and 9\%, using neutron
scattering measurements. \ The goal of this study is 1) to extend the
previous PZN study to other reciprocal lattice points , 2) to identify the
onset of the local polar order more definitely and verify its temperature
evolution and 3) to investigate the local structural and polar order at
higher PT ($x$) concentrations.

\section{Experiment}

Single crystals of PZN and PZN-9\%PT were grown by spontaneous nucleation
from high temperature solutions, using an optimized flux composition of PbO
and B$_{2}$O$_{3}$ \cite{Zhang etal_96_2000}. \ The PZN-4.5\%PT single
crystal was grown by the top-cooling solution growth technique, using PbO
flux \cite{Chen etal_4393_2001}. \ All as-grown crystals used in the
experiment exhibited a light yellow color and high optical quality.

The neutron experiments were carried out on the HB1 and HB1A triple-axis
spectrometers at the High Flux Isotope Reactor (HFIR) of Oak Ridge National
Laboratory and on the BT7 and BT9 triple-axis spectrometers at the NIST
Center for Neutron Research (NCNR). The neutron energies used for the data
displayed in this paper were 14.7 meV (2.36 \AA ) or 13.6 meV (2.45 \AA ) at
HFIR and 14.7 meV (2.36 \AA ), 13.4 meV (2.47 \AA ) or 30.4 meV (1.64 \AA )
at NIST. Highly oriented pyrolytic graphite (002) (HOPG) was used to
monochromate and analyze the incident and scattered neutron beams. An HOPG
filter was used to suppress harmonic contamination. \ The samples were
mounted on an aluminum sample holder, wrapped with copper foil and held in
place with either an aluminum or a copper wire. \ To prevent contamination
of the spectra by the scattering from aluminum, the sample holder was
painted with gadolinium oxide or boron nitride paste, then loaded into a
vacuum furnace. The measurements were made in the [100]-[011] scattering
plane, which allowed access to the [100], [011], and [111] symmetry
directions. \ Data were collected upon cooling from 650 K to 295 K around
several reciprocal lattice points. No external electric field was applied.

\section{Results and discussion}

%
%
\begin{figure}[tbp]
\includegraphics[height=0.6\textheight]{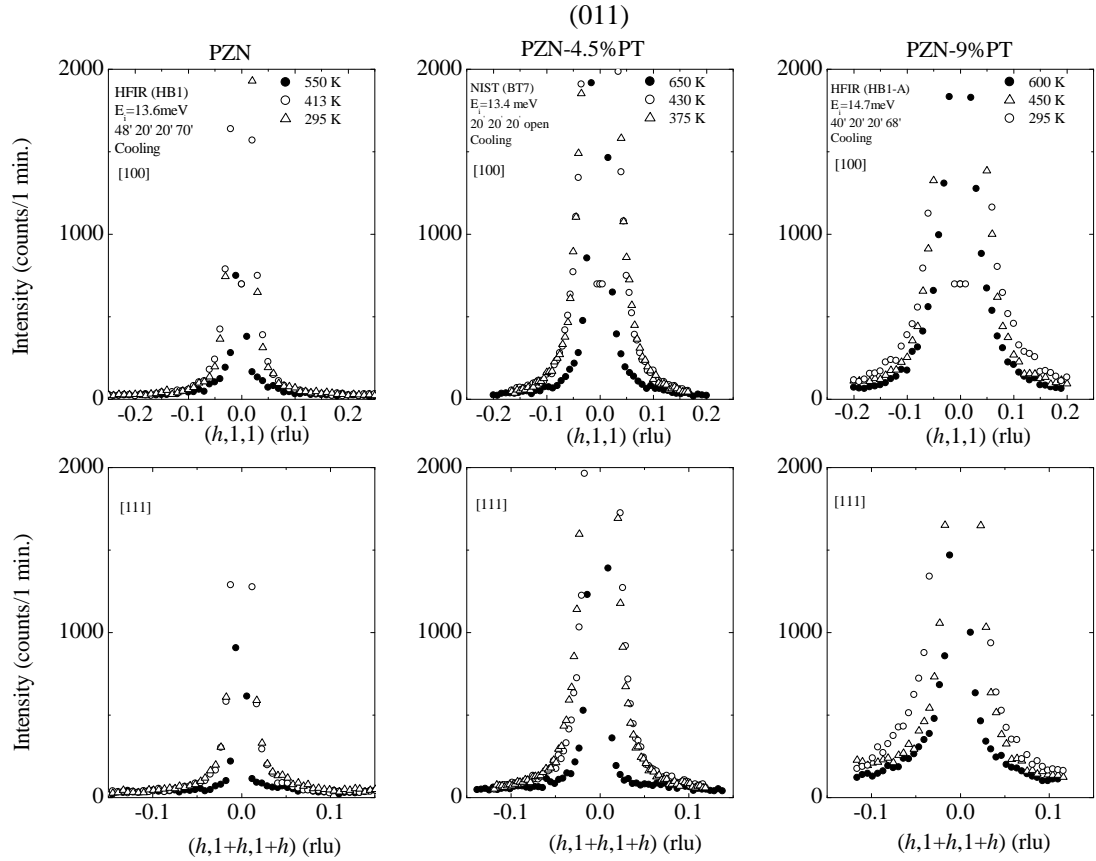}
\caption{Neutron elastic scattering profile for PZN, PZN-4.5\%PT, and
PZN-9\%PT at (011) point along [100] and [111] directions at different
temperatures, showing the narrow Bragg peak and the broad diffuse scattering
peak}
\label{fig1}
\end{figure}
%
%
The diffuse scattering measurements have been made in the [100]-[011] zone,
around many different reciprocal lattice points and in several directions.
The diffuse scattering is found only for points with mixed ($hkl$) indices.
In particular we have observed the diffuse scattering from (100), (011), and (300), 
which is consistent with the calculated diffuse
scattering intensity on PMN reported by Mathan {\it et al.}\cite{Mathan
etal_8159_1991} The diffuse scattering is observed along
transverse and diagonal directions only. No diffuse scattering is found
in a longitudinal direction. Like in many previous studies 
\cite{Vakhrushev etal_1995,Yong et al_14736_2000}, 
no diffuse scattering is observed at (200) reflection. 
This absence is either due to the soft mode, for which the structure factor 
is largest at the (200) in perovskites, or it must be related to the local 
order that develops below 600-700 K. Recent neutron diffuse scattering and 
the calculation model on PMN reported by Hirota {\it et al.}\cite{Hirota etal} 
have also confirmed the weak (200) diffuse scattering using a new concept of 
the phase-shifted condensed soft mode. A typical scan is presented
in Fig.~\ref{fig1}, which shows diffuse scattering around the (011) point
for PZN, PZN-4.5\%PT, and PZN-9\%PT single crystals at different
temperatures. With decreasing temperatures, the diffuse scattering peaks
become broader, more extended in the [100] direction in pure PZN and in the
[111] direction in PT-doped PZN. It also becomes broader with increasing PT
concentration. Below the transition, the width of the diffuse scattering
peaks remains constant, despite the fact that the Bragg peaks increase continuously. 
The diffuse scattering peaks are slightly asymmetric, which could be related to 
a rather large deformation of the crystal lattice giving rise to Huang 
Scattering \cite{Krivoglaz_NewYork_1969}.

%
%
\begin{figure}[tbp]
\includegraphics[width=0.6\linewidth]{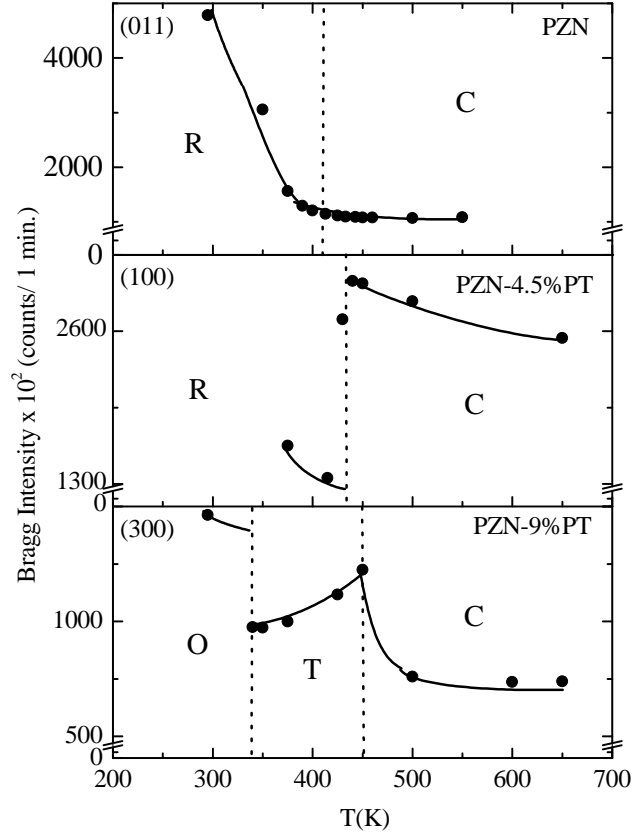}
\caption{Temperature dependence of Bragg intensities for PZN at (011),
PZN-4.5\%PT at (100), and PZN-9\%PT at (300) along the [111] directions,
showing the phase transitions. Solid lines are drawn through the data points as
guides to the eye. Dashed lines mark the phase transitions.}
\label{fig2}
\end{figure}
%
%

The temperature dependence of Bragg intensity presented in Fig.~\ref{fig2}
reveals the phase transition temperature $T_{c}$. It is seen that pure PZN and PZN-$x$PT
behave differently, the former undergoing a continuous
structural transition while the latter undergoes abrupt structural changes.
The rapid increase in Bragg intensity, observed for PZN below the transition $%
T_{c}$ $\sim $ 413 K, is due to the relief of extinction caused by a rapid
increase in mosaicity \cite{La-Orauttapong_etal_PRB}. In PZN-4.5\%PT, the
Bragg intensity increases moderately as the first transition is approached
but abruptly drops at $T_{c}$ $\sim $ 430 K, when the crystal structure
transforms from cubic to rhombohedral symmetry. \ In the 9\%PT crystal, the
first transition at 450 K is marked by a cusp while the second one at 340 K
is marked by a discontinuous increase. \ From these results, it is evident
that the addition of PT triggers sharp structural changes. This suggests that
PT-doping enhances the local strain field important for ferroelectric phase
transition. It is this strain field that induces the real phase transition
characterized by the long range structural order. The
transition temperatures observed in the present work are in good agreement
with those previously reported \cite{Kuwata_etal_579_1981}.

%
%
\begin{figure}[tbp]
\includegraphics[width=0.6\linewidth]{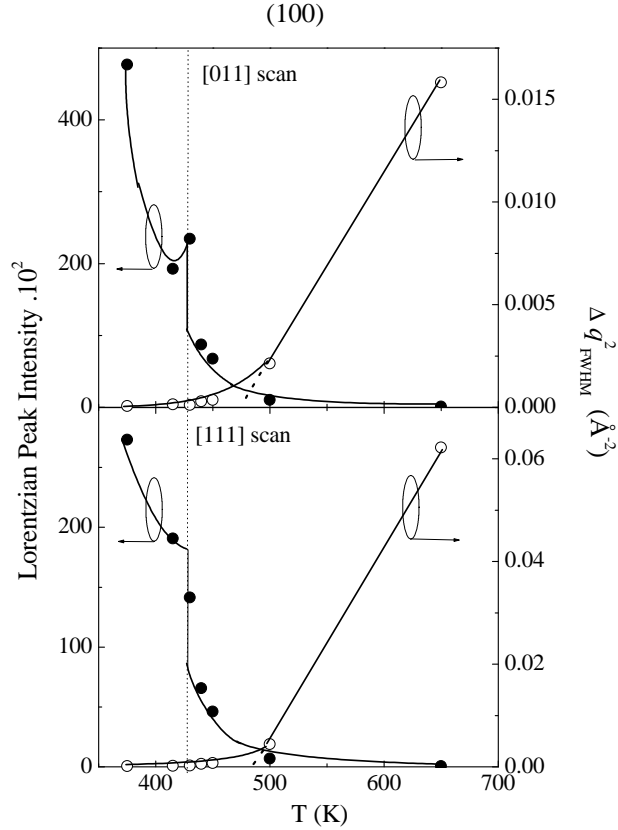}
\caption{Lorentzian Peak Intensity and square of the FWHM, $\Delta
q_{_{FWHM}}^{2}$ vs. temperature for PZN-4.5\%PT at the (100) point 
along the scan along the [011] and [111] directions.}
\label{fig3}
\end{figure}
%
%
%
%
\begin{figure}[tbp]
\includegraphics[width=0.45\linewidth]{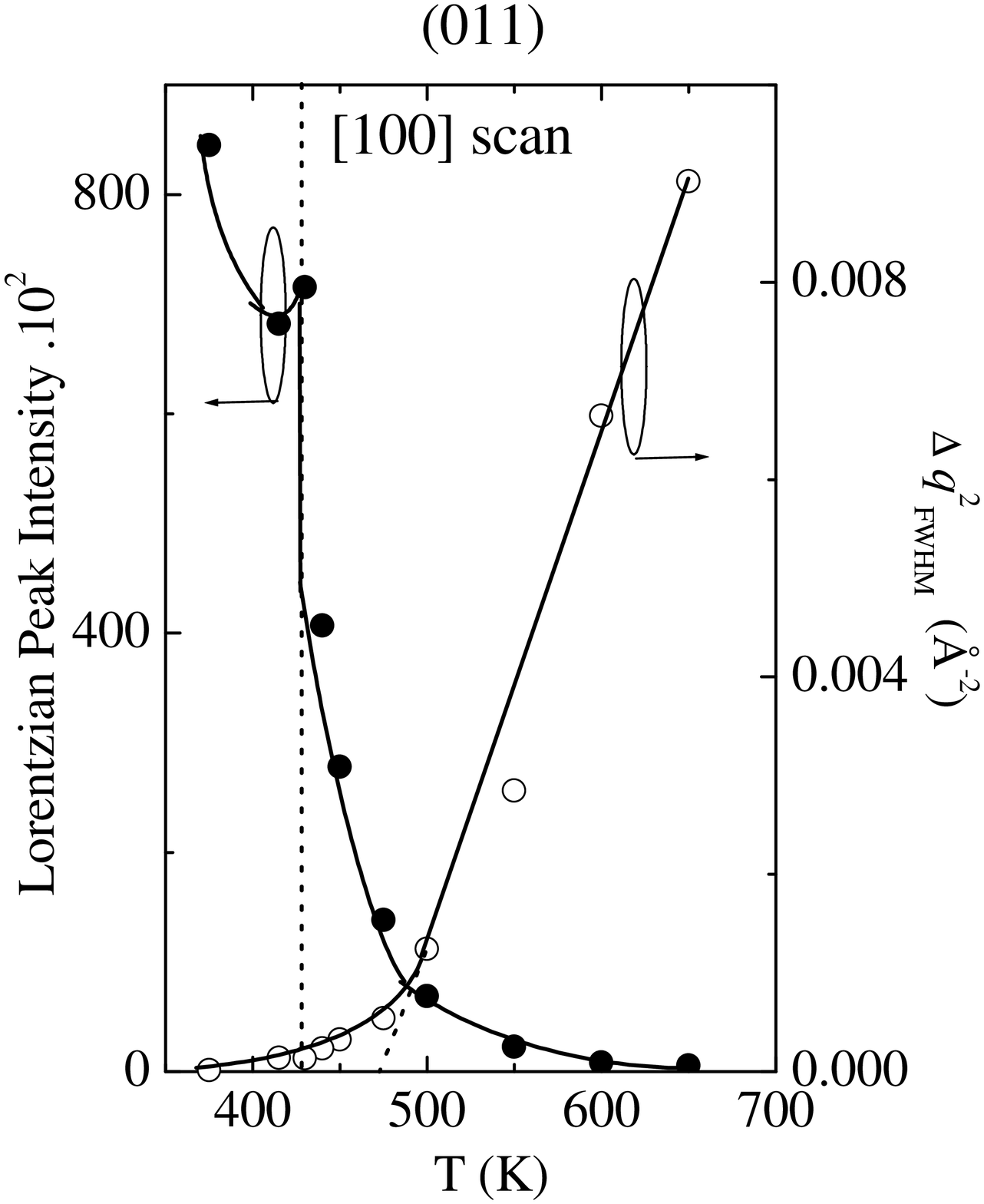}
\caption{Lorentzian Peak Intensity and square of the FWHM, $\Delta
q_{_{FWHM}}^{2}$ vs. temperature for PZN-4.5\%PT at the (011) point along
the [100] direction.}
\label{fig4}
\end{figure}
%
%

The observed diffuse scattering has been explained by the presence of polar
nanoregions, resulting from the short-range correlated atomic shifts \cite{Mathan
etal_8159_1991,Yong et al_14736_2000}. 
The reciprocal lattice points near which diffuse scattering is observed provide
information about the internal structure of the regions. The width of the
diffuse scattering gives us idea about the spatial extension of these regions in different
directions, i.e. their orientation.  In order to obtain information about
the size of the polar regions, we have fitted the spectra with a combination
of a Gaussian (Bragg) and a Lorentzian (diffuse), convoluted with the
experimental resolution function. A Lorentzian lineshape is predicted by the
Ornstein-Zernike model \cite{Stanley_1971}: 
\begin{equation}
I\simeq \frac{1}{q^{2}+\xi ^{-2}}
\end{equation}%
where $q$ $\equiv $ $h$ is the momentum transfer relative to the \textbf{Q}
= (011) Bragg reflection and $\xi $ is the correlation length. 
This model assumes a correlation function of the form: 
\begin{equation}
\frac{e^{-\frac{r}{\xi }}}{r}.
\end{equation}%
Therefore the diffuse scattering width at half
maximum, $\Delta q_{_{FWHM}}$, provides an estimate of $\xi $ or,
equivalently, of the size of the polar nanoregions: 
\begin{equation}
\xi =\frac{2}{\Delta q_{_{FWHM}}} ,  
\label{pnr}
\end{equation}%
where $\Delta q$ in Eq.(~\ref{pnr}) is in units of $\frac{2\pi }{a}$. The
diffuse scattering intensity is generally proportional to average polarization 
squared $\left\langle P_{local}^{2}\right\rangle$. An increase of the diffuse scattering
intensity with decreasing temperature reflects the growth of the polar
clusters or growing correlation of atomic displacements. \ Below a certain
temperature, $T_{f}$, the size of the polar regions stabilizes, as
indicated by a constant width of the diffuse scattering peak.  In the PZN
system, the strongest diffuse scattering is primarily associated with the
short-range correlated displacements of the Pb atoms along [111] direction 
\cite{La-Orauttapong_etal_PRB}. For PZN-4.5\%PT, the diffuse scattering becomes 
stronger in the transverse than in the [111] directions as shown on the left 
side of Fig.~\ref{fig3}. This figure presents the temperature dependence of 
the diffuse scattering in PZN-4.5\%PT
at the (100) reflection in the transverse and [111] directions. In both
directions, the diffuse scattering intensity increases continuously at
first, and then goes through a cusp at the transition. The cusp is less
pronounced in the [111] than in the transverse direction, looking more like
a discontinuity in slope. It is important to note that intensity in the transverse direction
is stronger than that in the [111] direction. Therefore, upon substitution 
of Ti$^{4+}$ for (Zn$_{1/3}^{2+}$/Nb$_{2/3}^{5+}$)$^{4+}$ at the B-site position, a change in symmetry of the
primary distortion occurs. 
%
%
\begin{figure}[tbp]
\includegraphics[width=0.45\linewidth]{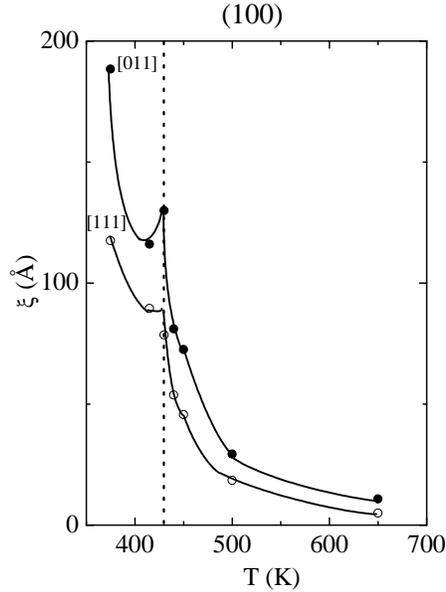}
\caption{Temperature dependence of the correlation length, $\protect\xi $
for PZN-4.5\%PT at the (100) reciprocal lattice point along the [011] and
[111] directions.}
\label{fig5}
\end{figure}
%
%
We have also found \cite{La-Orauttapong_etal_PRB} that at high temperature 
the correlation length squared $(\xi ^{2})$ follows a Curie-Weiss law. 
This suggests that the polar nanoregions are highly dynamics, 
since $\Delta q_{_{FWHM}}^{2}$ is inversely proportional to $\xi ^{2}$, which
is itself proportional to the dielectric constant, $\epsilon $: 
\begin{equation}
\Delta q_{_{FWHM}}^{2}\sim \frac{1}{\epsilon }=\frac{T-T_{c}}{C}
\end{equation}%
where $C$ is the Curie constant. Therefore, the deviation from the linear dependence 
confirms the appearance of long-lived polar fluctuations at $%
T^{\ast }=T_{c}+\delta T$ in the crystal, accompanied by local strain
fields. With PT addition, $\delta T$ increases from $\sim $
40 K (in PZN\cite{La-Orauttapong_etal_PRB}) to $\sim $ 70 K (in PZN-4.5\%PT
as shown on the right side of Fig.~\ref{fig3} and ~\ref{fig4}). It means the process of slowing 
down in PZN-$x$PT begins at higher temperature than in pure PZN. 
This fact can be ascribed to the enhanced correlations between
polar clusters induced by the presence of PT. This is in agreement to the similar results on PZN and
PZN-10\%PT single crystals deduced from the Vogel-Fulcher relation 
by Seo {\it et al} \cite{Seo etal_496_1999}. They have shown that the static freezing 
temperature ($T_{f}$) in PZN ($T_{f}$=240 K) is lower than that in PZN-10\%PT ($T_{f}$=297 K). 

In Fig.~\ref{fig5}, we present the temperature dependence of the correlation
length for PZN-4.5\%PT at the (100) reflection along the transverse and
diagonal directions. At the temperature $T^{\ast }$, the size of these
regions is finite, about 29 \AA ($\sim$7 unit cells) and 18 \AA ($\sim$4 unit cells)
along the transverse and diagonal directions, respectively. As the
temperature decreases, the correlation lengths increase continuously at first, and
then go through a cusp at the transition. In phase transition phenomena,
cusps often indicate the occurence of a condensation (see e.g. the
liquid-gas transition, with a possible supercooled vapor phase). 

In this study we have shown that the polar regions in PZN are preferentially more 
extended in the [111] than in the transverse direction, which is consistent with the 
[111] displacement of the Pb atom. However, with the addition of PT, the preferred orientation of the
regions switches to the transverse direction. Nonetheless, the present
diffuse neutron scattering results suggest that, upon going from PZN to PZN-$%
x$PT, the local polarizations (i.e. in the polar nanoregions) still retain an
internal rhombohedral symmetry. This is suggested by the fact that, so far,
the diffuse scattering  has been observed at the same reciprocal points
(100), (011), and, (300) in PZN and PZN-$x$PT.  If this observation is
confirmed, this would mean that, with increasing PT, the polarization
direction does not rotate, but that the regions preferentially grow  in a
different direction. Expressed differently, with the addition of PT, the
polarization is still pointing in the [111] direction but the polar regions
become longer in the transverse than in the [111] direction. This should
have a positive effect on the ease of reorientation of this polarization,
making it easier in PZN-$x$PT than in pure PZN. Before this point can be
settled, additional measurements need to be performed in a different
scattering zone.

\begin{theacknowledgments}
 This research has been supported by DOE under Contract No. DE-FG02-00ER45842 for the experimental 
 part and by ONR under Grant No. N00014-99-1-0738 for the crystal growth. We acknowledge the support 
 of NIST Center for Neutron Research and Oak Ridge National Laboratory, in providing the neutron 
 facilities used in this work and also thank R.K. Pattnaik, and O. Svitelskiy for their helpful 
 suggestions.  
\end{theacknowledgments}

\bibliographystyle{aipproc}   


\end{document}